\documentclass[aps,prl,twocolumn,superscriptaddress,showpacs]{revtex4}
\usepackage{graphicx}
\usepackage{amsmath}
\usepackage{natbib}
\begin{document}
\title{Different Time Scales in Wave Function Intensity Statistics}
\author{D. A. Wisniacki}
\affiliation{Departamento de Qu\'{\i}mica C--IX,
 Universidad Aut\'onoma de Madrid,
 Cantoblanco, 28049--Madrid (Spain).}
\affiliation{Departamento de F\'{\i}sica,
 Comisi\'on Nacional de Energ\'{\i}a At\'omica.
 Av.\ del Libertador 8250, 1429 Buenos Aires (Argentina).}
\author{F. Borondo}
\email[Corresponding author: ]{f.borondo@uam.es}
\affiliation{Departamento de Qu\'{\i}mica C--IX,
 Universidad Aut\'onoma de Madrid,
 Cantoblanco, 28049--Madrid (Spain).}
\author{E. Vergini}
\affiliation{Departamento de F\'{\i}sica,
 Comisi\'on Nacional de Energ\'{\i}a At\'omica.
 Av.\ del Libertador 8250, 1429 Buenos Aires (Argentina).}
\author{R. M. Benito}
\affiliation{Departamento de F\'{\i}sica,
 E.T.S.I.\ Agr\'onomos, Universidad Polit\'ecnica de Madrid,
 28040 Madrid (Spain).}
\date{\today}
\begin{abstract}
Unstable periodic orbits scar wave functions in chaotic systems.
This also influences the associated spectra, that follow the
otherwise universal Porter--Thomas intensity distribution.
We show here how this deviation extend to other longer periodic
orbits sharing some common dynamical characteristics.
This indicates that the quantum mechanics of the system can be
described quite simply with few orbits, up to the resolution
associated to the corresponding length.
\end{abstract}
\maketitle

Chaos is a well defined phenomenon in classical mechanics, but its
manifestations at quantum level are not fully understood yet \cite{1}.
The pioneering work of Gutzwiller \cite{gut},
showing how the eigenvalues density of chaotic systems can be obtained
from classical periodic orbits (PO), constitutes an important landmark
in the investigation of ``quantum chaos''.
Later Heller demonstrated that, contrary to reasonable conjectures,
unstable POs have a profound influence in
the distribution of quantum probability density of a non--vanishing
class of wave functions, which appear highly localized over these
classical paths \cite{Heller2}.
The wave functions exhibiting this localization are said to be
``scarred'', and they play a very important role in semiclassical
theories \cite{Heller3}.
Recently, several methods, based on very different strategies,
have been described in the literature \cite{pol,ver2}
for the systematic construction of non--stationary wave functions
highly localized on POs.
These novel tools also provide new insight towards the understanding
of the role of scarring in the quantum mechanics of chaotic systems
\cite{ver3}.
Scars also condition spectra.
Recurrences in the correlation function originated by the linearized
dynamics around unstable POs are the reason for the existence
of marked low resolution structure in the spectra of chaotic systems.
Heteroclinic orbits have also been demonstrated to be important
when extending this theory beyond the linearized regime
\cite{Heller4,Tomsovic}.

Another main achievement in quantum chaos is undoubtedly Random Matrix
Theory (RMT) \cite{RMT}, which accounts for many properties in the
quantum spectra of chaotic systems, such as the widespread nearest
neighbor energy level spacing, which for all strong mixing systems
follows the Wigner surmise \cite{Bohigas}.
The beauty of many RMT results is their universality, which turns out
to be also its drawback, since they are independent of the initial
state preparation details and its dynamics.
Other measures sensitive to them, such as the distribution of spectral
intensities, seem then, in principle, better suited to elucidate
relevant features in a given spectrum.
However, the statistical fluctuations of quantum transition strengths
in stochastic systems was also found to be described by another RMT
universal formula, the Porter--Thomas (PT) distribution \cite{PT}.
In this respect, Alhassid and Levine demonstrated \cite{Alhassid}
that the PT distribution can be simply obtained from the principle of
maximum entropy of the strength distribution, with the only constrain
of the ever present sum rule for the total strength of the transition.
Sibert and Borondo \cite{Sibert} showed that the same
result can be derived by imposing the dynamical constrains
implied by the short time motion of the system.
Later Kaplan \cite{Kaplan} found that,
although the tail of wave function intensity distribution in phase
space is dominated by scarring associated with the least unstable PO,
when the low resolution modulation induced by it is removed,
the remaining distribution matchs the standard PT expression.

In this Letter we report an study on the intensity distribution
statistics for the stadium scar wave functions,
calculated with the method of Ref.~\onlinecite{pol}.
We show that the information contained in it about POs is far richer than
assumed in the previously existing literature \cite{RMT,Sibert,Kaplan},
actually extending to times much longer than that of the local short
term dynamics dictated by the least unstable PO.
However, this view is based on few POs, thus representing an
important conceptual simplification over other approaches
(see for example \cite{Heller4,Tomsovic}), which can be very useful
in future developments of scar theory.

In our study we use a system consisting of a particle of mass 1/2
enclosed in a desymmetrized stadium billiard of radius $r=1$ and area
of $1+\pi/4$, with Newman boundary conditions on the symmetry axes
(only even--even parity eigenfunctions will be considered).

We consider the dynamics influenced by the horizontal PO running along
the $x$ axis with $y=0$.
For this purpose we start from a symmetry adapted initial wave packet:
\begin{equation}
 \begin{array}{lcl}
   \langle x,y|\phi(0)\rangle & = & a G_{x^0,y^0,P_x^0,P_y^0}
       +   b G_{-x^0,y^0,-P_x^0,P_y^0}  \\
     & + & c G_{x^0,-y^0,P_x^0,-P_y^0} \\
     & + & d G_{-x^0,-y^0,-P_x^0,-P_y^0} + \text{c.c.}
 \end{array}
 \label{eq:phi0}
\end{equation}
where
\begin{equation}
  G_{{\bf q^0},{\bf P^0}} = \displaystyle\prod_j
         (\pi\alpha_j^2)^{-1/4}
         {\text{e}}^{-(q_j-q_j^0)^2/2\alpha_j^2} \;
         {\text{e}}^{\text{i}P_j^0(q_j-q_j^0)},
 \label{eq:G}
\end{equation}
(the coefficients $a,b,c$, and $d$ are obtained by imposing Newman
boundary conditions at the symmetry axes), and compute the
(infinite resolution) spectrum
\begin{equation}
  I_\infty(E) = \sum_n |\langle n|\phi(0)\rangle|^2 \delta(E-E_n),
 \label{eq:I(E)}
\end{equation}
with $(x^0,y^0,P_x^0,P_y^0)$=$(1,0,k_0,0)$ and
$\alpha_x$=$\alpha_y$=$1.603/k_0^{1/2}$ ($\hbar$ is set equal to 1
throughout this paper).
Kets $|n\rangle$ represent the eigenstates of the system, which have
been obtained using the scaling method \cite{sca}.
The aspect of the obtained $I_\infty(E)$ is rather irregular
(see results below),
due to the highly chaotic nature of the dynamics in this system.
However, when examined closely, the contributing ``sticks'' are seen
to come grouped in clumps.
This indicates the existence of a clear underlying structure,
which is due to recurrences in the associated correlation function,
$C(t)=\langle \phi(0)|\phi(t)\rangle$, induced by the horizontal PO
used to select the initial position of the wave packet.
This regularity shows up as well defined bands in the low resolution
version of the spectrum, $I_T(E)$,
\begin{equation}
  I_T (E) = \frac{1}{2\pi}\int_{-\infty}^\infty \; dt \;
     W_T (t) C(t) \exp(iEt),
 \label{eq:I_T(E)}
\end{equation}
where $W_T (t)$ is a suitable smoothing window function, filtering
out the dynamics of the system for times longer than $T$.
Moreover, the positions of these bands are given by the usual
Bohr--Sommerfeld quantization condition,
\begin{equation}
  k_m=\frac{2\pi}{L_\mu}\left(m+\frac{\nu_\mu}{4}\right),
 \label{eq:BS}
\end{equation}
with $L_\mu$=4 and $\nu_\mu$=3.
The wave functions associated to these bands correspond to a series of
scar functions on the PO with an increasing excitation along it,
as discussed elsewhere \cite{pol,Diego1,Diego2}.

\begin{figure}[tb]
 \includegraphics[width=7.5cm]{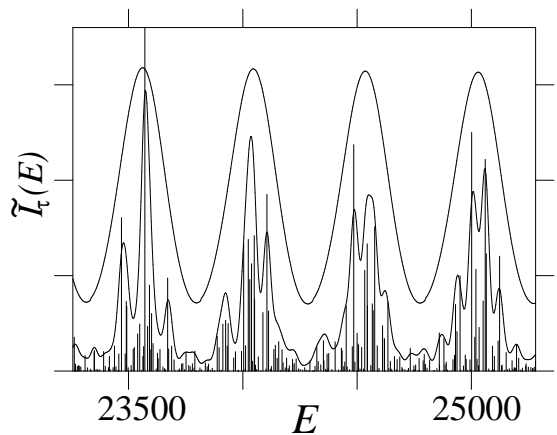}
 \caption{\label{fig:1}
   Scar spectrum, ${\tilde I}_\tau(E)$, at infinite (stick) and
   low resolution: $\tau=1$ and 4.5 (solid lines),
   corresponding to a wave packet initially centered on the horizontal
   axis of a desymmetrized stadium billiard with $r=1$, area $1+\pi/4$,
   and Newman boundary conditions at the symmetry axes.}
\end{figure}

To efficiently study the characteristics of the spectra corresponding
to this band structure we define a ``scar spectrum'' in the
following way.
Using the procedure described above, we calculate spectra at all
energies, $k_0$=$k_m$, quantized with the Bohr--Sommerfeld formulae
and construct a new spectrum, ${\tilde I}_\infty (E)$, by taking
only the central clump from each one of them:
\begin{equation}
   {\tilde I}_\infty(E) = \sum_m I_m^{\text{band}}
   = \sum_m \sum_{\{E_n\}}{'}|a_n^m|^2 \; \delta(E-E_n),
 \label{eq:Itilde1}
\end{equation}
where the prime indicates that the sum is only carried out for states
in the range ($E_m$--$E_{m-1}$)/2$<$$E_n$$<$($E_{m+1}$--$E_m$)/2.
In this way the spectrum statistics is improved, since the constrain
imposed by the finite width of the initial wave packet is eliminated.
Our calculations were performed in the range $k$=50--250,
which includes approximately 8400 stadium eigenstates.
Similarly to what was done in eq.~(\ref{eq:I_T(E)}), we can now define
a low resolution version of this spectrum which, for a Gaussian window
function, takes the form
\begin{equation}
  {\tilde I}_\tau(E) = 2 \tau T_0 \pi^{-1/2}
    \sum_{\{E_n\}}{'} |a_n|^2 \; {\text{e}}^{-2\tau^2 T_0^2 (E-E_n)^2},
 \label{eq:Itilde2}
\end{equation}
where $\tau$ is an adimensional smoothing parameter and $T_0$ the
period of the scarring orbit.
The corresponding results, in the range $E$=23250--25300,
are shown in Fig.~\ref{fig:1}, for $\tau$=1 and $4.5$.
The first value of the smoothing parameter corresponds approximately
to the smallest smoothing which washes out all substructures.
The required time scales with the inverse of the Lyapunov exponent
\cite{Heller2}.
When the resolution is increased to $\tau$=4.5 another superimposed
intra--band structure is then exposed, thus revealing the relevance
of longer time dynamics.

%

To quantitatively examine the implication of this result, we consider
next the statistics of intensities in our scar spectrum.
According to RMT the distribution of ``dynamically normalized''
intensities in a fully chaotic system is given by the usual
$\chi^2$ (PT) fluctuations \cite{PT},
\begin{equation}
    P(x) = (2\pi x)^{-1/2} {\text{e}}^{-x/2}.
 \label{eq:PT}
\end{equation}
However, and as stated in the introduction, one crucial point when
performing this analysis is to eliminate the modulation due to any
obvious low resolution structure present in the spectrum, so that all
intensities are compared on the same relative scale.
This can be acomplished by ``renormalizing'' the intensities with
the corresponding value of the envelope \cite{Sibert,Kaplan},
\begin{equation}
   x_n = N |a_n|^2/{\tilde I}_\tau(E_n),
 \label{eq:xn}
\end{equation}
where, coherently with eq.~(\ref{eq:PT}), the number of states in the
spectrum, $N$, has been included, in order to obtain a mean value
of unity.
In Ref.~\onlinecite{Sibert} it was shown that this task can be performed
systematically by monitoring the corresponding variance
from the PT distribution, $\sigma^2$, as a function of the
smoothing time.

%
\begin{figure}[tb]
 \includegraphics[width=6.5cm]{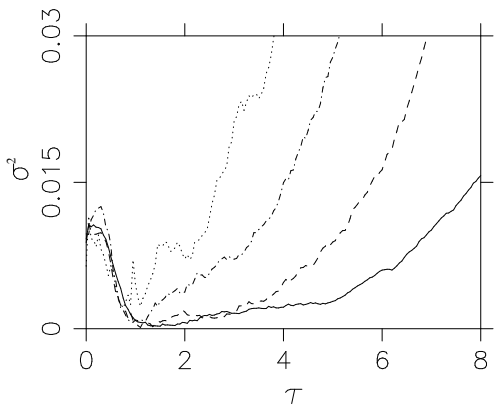}
 \caption{\label{fig:2}
   Variance from the Porter--Thomas results of the
   wave function intensity distribution for the scar spectrum
   of Fig.\protect\ref{fig:1}, calculated using the first 1000 (dotted),
   2000 (dotted--dashed), 4000 (dashed), and 8000 (full line) states
   of the scar spectrum of Fig.~\protect\ref{fig:1}.}
\end{figure}

In Fig.~\ref{fig:2} we show the results of this calculation when using
the first 1000, 2000, 4000, and 8000 states of the scar spectrum of
Fig.~\ref{fig:1}.
In order to eliminate the contribution from the tail of the
distribution \cite{Kaplan} and the divergence at the origin of the PT
expression, only values in the range 1$<$$x_n$$<$5 have been included
in our statistical treatment.
As can be seen, all curves fall off initially quite rapidly as $\tau$
increases, and then start to stabililize at $\tau$$\simeq$1, indicating
that, at least, an optimal fitting to the PT surmise can be obtained
when substracting from the intensity distribution the low resolution
envelope corresponding to the dynamics of the horizontal PO.
This result is in agreement with the conclusions of \cite{Kaplan}.
It is worth to emphasize that the value of $\tau$ for which the PT
statistics starts to work is not directly related to the period of
the orbit.
For instance, for an orbit with a very small Lyapunov exponent,
the value of $\tau$ would be much greater than 1.
More interesting is that, after this point, more optimal values of
$\sigma^2$ are obtained for a range of values of $\tau$, which extend
in some sort of plateau or broad minimum.
Finally, the variance grows (approximately) parabolically after a
point $\tau_{\text{max}}$, indicating when we are trying to
describe the spectrum with too much ``dynamical'' resolution
(for the number of states that have been included in the statistics).
The relevance of this figure is that it shows the existence of plateaus,
whose extension grow with $N$.
Actually, $\tau_{\text{max}}$ is found to scale, similarly to the
Heisenberg time, with $\sqrt{N}$, although it differs from this
expression by a factor of $\simeq$10, indicating that our statistical
treatment should allow at least 10 states per band in order to capture
the fluctuations implied by the PT distribution.

\begin{figure*}[tb]
 \includegraphics[width=12.0cm]{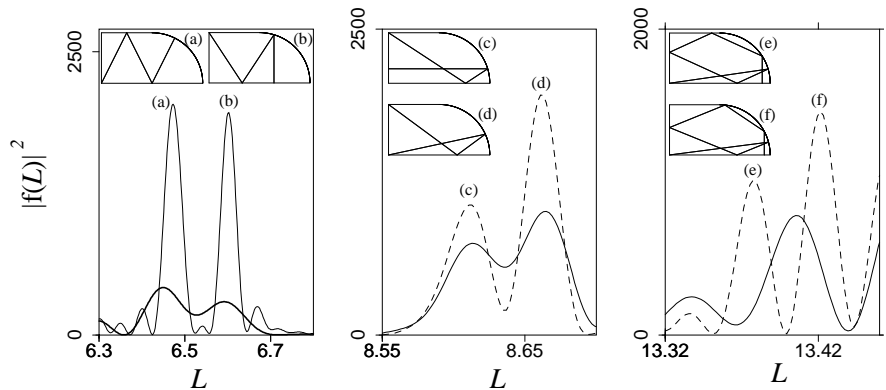}
 \caption{\label{fig:3}$|f(L)|^2$ computed from the eigenvalues
  density of the desymmetrized stadium with Newman boundary conditions
  for $N$=1000 (thick solid line), 2000 (thin solid line),
  and 8000 (dashed line), and the corresponding periodic orbits.}
\end{figure*}

This result is further illustrated in Fig.~\ref{fig:3}, where,
in addition to that, the relation between $L_{\text{max}}$,
the corresponding orbit length, and invariant classical structures
is also revealed.
To interpret this figure one must take into account that in the
stadium, as in any other chaotic system, Gutzwiller's trace formulae,
$\rho(k)=\sum_n \delta(k-k_n)$, gives the quantal density of states
in terms of information on all POs of the system.
This process can be inverted, by Fourier transform,
to obtain the classical spectra of orbits, $f(L)=\sum_n \exp(ik_n L)$.
Figure \ref{fig:3} shows the square of this magnitude computed for
$N$=1000, 4000 and 8000.
The results in the first panel indicate that 1000 states are barely
enough to distinguish between the two PO of length $L$$\simeq$6.5
plotted in the insets, which on the other hand are fully resolved
when $N$=4000.
The central panel shows how 4000 states are able to discern dynamical
features up to $L$$\simeq$9.
And finally, the results of panel three imply that the quantum mechanics
of the system up to $L$$\simeq$16 can be described with 8000 states.

%

\begin{figure}[tb]
 \includegraphics[width=6.5cm,angle=-90]{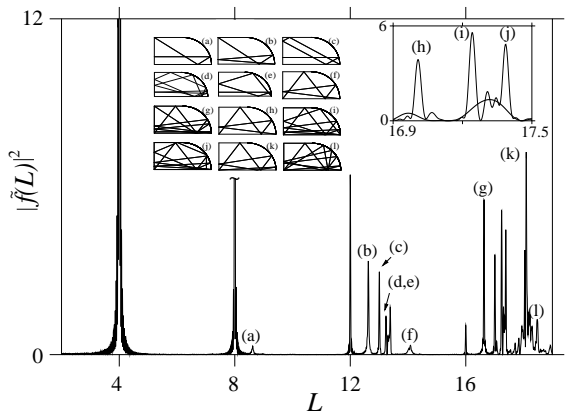}
 \caption{\label{fig:4}
   $|{\tilde f}(L)|^2$ computed from the stick scar spectrum of
   Fig.~\protect\ref{fig:1}.
   The different peaks has been assigned to the periodic orbits
   plotted in the left inset.
   The right inset shows how 1000 states (thick solid line) is not
   enough to reproduce the three central peaks of the fourth
   recurrence in $|{\tilde f}(L)|^2$.}
\end{figure}

In order to elucidate which POs, other than the original horizontal one,
are responsible for the plateaus observed in Fig.~\ref{fig:2},
a similar analysis can be performed with the scar spectrum of
Fig.~\ref{fig:1}, by using the following strategy.
Instead of the global density of states, $\rho(k)$, we employ now the
local (around the horizontal PO in phase space) version of it,
\begin{equation}
   {\tilde f}(L) = \sum_n |a_n|^2 \; {\text{e}}^{i k_n L}.
 \label{eq:f(L)}
\end{equation}
The inclusion of the weights $|a_n|^2$ it is not an irrelevant point,
since it implies that, contrary to what happens in Gutzwiller's
original trace formula, only POs dynamically linked to the initial
one are allowed to enter in our calculations \cite{Diego2}.
The corresponding result is presented in Fig.~\ref{fig:4},
where it is seen that $|{\tilde f}(L)|^2$ presents a series of prominent
peaks at multiples of a fundamental length of 4, the length of the
horizontal PO along which the packet was initially launched.
Moreover, the contribution of other, longer POs is also clearly observed.
By considering the lengths of the different POs of the stadium, we have
been able to assign each of the (non--trivial) contributing peaks,
up to the fourth recurrence of the horizontal orbit;
the corresponding POs are presented in the left inset of the figure.
Notice that all these POs present some good portion ot their paths
overlaping significantly with the initial horizontal PO.
Again the degree of resolution of our calculation is related to the
maximum value of $k$ included in the spectrum.
This is illustrated in the upper right inset to Fig.~\ref{fig:4},
where a blown up of the fourth recurrence in $|{\tilde f}(L)|^2$,
calculated using 1000 and 8000 states of the scar spectrum of
Fig.~\ref{fig:1}, is presented.

%

One final point, worth mentioning, is the heights of the peaks in
Fig.~\ref{fig:4}.
Since they contain information on the degree and phase of the
interaction between the different POs contributing to
$|{\tilde f}(L)|^2$, it must be possible to obtain from them
some clue of the scarring process beyond the short time limit
corresponding to the linearized dynamics around the initial
unstable fixed point.
This interaction can be evaluated, for example, as the Hamiltonian
matrix elements of the scar wave functions obtained with the methods
of Refs.~\onlinecite{pol,ver2,ver3}, and this will be the subject of
a future publication.

In summary, we have shown how scar spectra, containing information
about POs dynamically related, can be constructed for highly chaotic
systems.
This information is revealed by analyzing the associated intensity
distributions, which show, superimposed to the universal PT behavior,
low resolution structures in the range of time scales of the
corresponding periods.
\begin{acknowledgments}

This work was supported by BMF2000--437 DGES, AECI (Spain),
PICT97 3--50--1015, SECYT--ECOS and CONICET (Argentina).
\end{acknowledgments}
%

%
\end{document}